# Investigation of Thermodynamic Properties of Cu(NH$_3$)$_4$SO$_4$·H$_2$O, a Heisenberg Spin Chain Compound

**Tanmoy Chakraborty,** [1, 2] **Harkirat Singh,** [1, 3] **Dipanjan Chaudhuri,** [1, 4] **Hirale S. Jeevan,** [5] **Philipp Gegenwart,** [5,6] **Chiranjib Mitra.** [1]

[1] Indian Institute of Science Education and Research (IISER) Kolkata, Mohanpur Campus, PO: BCKV CampusMain Office, Mohanpur - 741252, Nadia, West Bengal, India

[2] Fakultät Physik, Technische Universität Dortmund, D-44221 Dortmund, Germany

[3] Tata Institute of Fundamental Research, Homi Bhabha Road, Colaba, Mumbai 400 005, India

[4] The Institute for Quantum Matter, Department of Physics and Astronomy, The Johns Hopkins University, Baltimore, MD 21218 USA

[5] I. Physikalisches Institut, Georg-August-Universitat, D-37073 Göttingen, Germany

[6] Experimental Physics VI, Center for Electronic Correlations and Magnetism, University of Augsburg, 86159 Augsburg, Germany

E-mail: chiranjib@iiserkol.ac.in

**Abstract**

Detailed experimental investigations of thermal and magnetic properties are presented for Cu(NH$_3$)$_4$SO$_4$·H$_2$O, an ideal uniform Heisenberg spin ½ chain compound. A comparison of these properties with relevant spin models is also presented. The temperature dependent magnetic susceptibility and specific heat data has been compared with the exact solution for uniform Heisenberg chain model derived by means of Bethe ansatz technique. Field dependent isothermal magnetization curves are simulated by Quantum Monte Carlo technique and compared with the corresponding experimental ones. Specific heat as a function of magnetic field (up to 7T) and temperature (down to 2K) is reported. Subsequently, the data are compared with the corresponding theoretical curves for the infinite Heisenberg spin ½ chain model with J=6K. Moreover, internal energy and entropy are calculated by analyzing the experimental specific heat data. Magnetic field and temperature dependent behavior of entropy and internal energy are in good agreement with the theoretical predictions.

1. Introduction

In recent times, numerous successful efforts have been devoted to explore spin systems with low dimensional magnetic interactions. Many low dimensional spin systems could be suitably described by physical models there by capturing the essential experimental manifestations [1-6]. For instance, dimerized and uniform spin ½ chain, spin ladder, frustrated spin systems etc. are some of the well studied models where the theoretical predictions have been verified experimentally [2, 4 and 5]. Theoretical descriptions of low dimensional spin systems have been realized using powerful numerical techniques like Quantum Monte Carlo (QMC) simulations [7, 8], Exact Diagonalization (ED) method [9] and Transfer Matrix Renormalization Group (TMRG) [10] on one hand, whereas on the other hand, analytical methods like field-theoretical approaches [11] and Bethe ansatz [12] have been successful in calculating various thermodynamic properties that have been predicted theoretically. Successful experimental preparation of materials which are representative of these different spin models allows researchers to investigate various thermodynamic quantities using proper theoretical tools. Thus a fruitful connection has been established between theory and experiment [1-6 and 13]. Low dimensional spin systems have attracted attention of an immense number of researchers, mainly due to the fact that the ground states of these systems show some unique features. For instance, the ground state of a Heisenberg spin chain is entangled which contributes to non-zero entanglement even at finite temperatures owing to a weighted thermal mixture of ground and excited states [14]. Thus, one can experimentally capture the existence of entanglement in the thermal states of a given system. Low dimensional spin systems provide an opportunity to study a wide range of physical properties due to their exotic nature [2]. As illustrations, investigation of spin-gap excitation using far-infrared spectroscopy [15], study of thermal and spin transport properties at finite temperature [16], observation of magnetic singlet bound states using light scattering experiments [17] etc. have been performed on these systems. However, choosing magnetic materials, whose behaviors resemble predicted theoretical models, have indeed opened the window of investigation of these exotic features extensively.

Low dimensional quantum spin ½ systems exhibit novel magnetic and thermal properties. A well-known Spin-Peierls compound $CuGeO_3$, which possesses a dimerized spin state below a critical temperature $T_c$=14K, has shown direct evidence for a singlet-triplet transition in neutron inelastic scattering experiment [18, 19]. Another well studied spin ½ system τ-$(BEDT-TTF)_2Cu_2(CN)_3$, which exemplifies an organic Mott insulator with frustrated quantum spins, could be best modeled by a triangular-lattice Heisenberg model with exchange coupling constant $J \sim 250K$ as corroborated by $^1H$ NMR and static susceptibility measurements [20]. $TlCuCl_3$, $BaCuSi_2O_6$ and $Cu(NO_3)_2$-$2.5D_2O$ are three extensively studied dimerized quantum antiferromagnets where quantum phase transition and Bose-Einstein condensation of magnons have been experimentally observed ([21] and references therein). $SrCu_2(BO_3)_2$ is another widely investigated compound showing evidence for localized singlet-triplet excitation and is a representative of the Shastry-Sutherland model of a 2D spin-gap system with an exact dimer ground state [22]. The class of spin ½ Heisenberg chain materials belonging to the family of low dimensional spin systems, has fascinated many researchers in last few decades. $CuSe_2O_5$ [23] and $KCuGaF_6$ [24] are two good examples of spin chain materials which have demonstrated promising experimental results and a striking match with theoretical calculations. $Sr_2CuO_3$, which has recently been reported to show Spin-Orbital separation [25], is another good example of 1D isotropic antiferromagnetic spin chain with extremely large exchange coupling constant (~ 1300K). Furthermore, temperature and field dependent studies of magnetic and thermal properties have been reported for $Cu(C_4H_4N_2)(NO_3)_2$ and $[Cu(\mu-C_2O_4)(4-aminopyridine)_2(H_2O)]_n$ which are also good realizations of spin ½ Heisenberg chain [26, 27].

In the present work, we have experimentally investigated detailed magnetic and thermal properties of such an antiferromagnetic uniform spin ½ chain with isotropic Heisenberg interaction. The Hamiltonian of a one dimensional antiferromagnetic spin chain with S=1/2 in presence of applied external magnetic field can be written as,

$$H = \frac{J}{4}\sum_i \left[\alpha\sigma_i^z\sigma_{i+1}^z + \beta\left(\sigma_i^x\sigma_{i+1}^x + \sigma_i^y\sigma_{i+1}^y\right)\right] + g\mu_B B\sum_i \sigma_i^z \tag{1}$$

Where $0 \leq \alpha \leq 1$, $0 \leq \beta \leq 1$, $B$ is the external magnetic field, $\sigma^x$, $\sigma^y$, $\sigma^z$ are the three Pauli spin matrices, $g$ is the Landé g factor, $\mu_B$ is the Bohr Magneton and $J$ is the exchange coupling constant. The summation is taken over nearest neighboring spins. The system we have studied in the present work, is $Cu(NH_3)_4SO_4 \cdot H_2O$ which could be described by isotropic Heisenberg model with $\alpha = \beta = 1$ in equation (1) [28, 29]. Jong *et al.* have described the early experimental results for the present compound and established it as a good realization of exchange coupled Heisenberg spin chain [4]. Zero field magnetic susceptibility as a function of temperature has shown the existence of a broad maximum around 3.5K whereas the temperature dependent specific heat curve has shown a peak around 3K [4, 28 and 29]. These thermal and magnetic measurements have ascertained the existence of magnetic interaction along a particular direction in $Cu(NH_3)_4SO_4 \cdot H_2O$ [4, 28 and 29]. The crystallographic structural analysis reported by Fierenzo Mazzi has revealed the spin chain behavior in $Cu(NH_3)_4SO_4 \cdot H_2O$ [30]. In a linear chain, $Cu^{++}$ ions are connected along the c axis as — $Cu^{++}$—$H_2O$— $Cu^{++}$— $H_2O$— while the $Cu^{++}$ ions of neighboring chains are connected as — $Cu^{++}$—$NH_3$—$SO_4$—$NH_3$— $Cu^{++}$—, which results in larger intrachain coupling strength in comparison to that between the chains. Interaction with next nearest neighboring spins along the chain is also negligible here.

We have made a detailed estimation of various thermodynamic quantities from experimentally obtained data. The data has been analyzed within the framework of Heisenberg antiferomagnetic spin chain model with full isotropy. Temperature dependence of magnetic susceptibility and field dependence of isothermal magnetization are investigated experimentally. Furthermore, the heat capacity data were collected both at zero field and in presence of externally applied magnetic field. Exact results for Heisenberg spin chain have been compared with the experimental data. Subsequently, zero field and field dependence specific heat data were used to obtain experimental estimates of internal energy and entropy.

2. **Experimental Details**

The experiments were performed on single crystalline $Cu(NH_3)_4SO_4 \cdot H_2O$ of purest grade (99.999 %), supplied by Sigma Aldrich. Magnetic susceptibility and magnetization measurements were performed in a Quantum Design Magnetic Property Measurement System (MPMS) and Oxford Vibrating Sample Magnetometer (VSM). Temperature dependence of static magnetic susceptibility was measured in a temperature range of 1.9 K to 30 K. Subsequently, isothermal magnetization measurements were carried out as a function of field at various temperatures. The field was varied from 0T to 14 T and the temperature was varied from 1.9 K to 10 K.

The specific heat measurements were performed by a standard relaxation method in a Quantum Design Physical Property Measurement System (QD PPMS). Zero field specific heat data were collected in a temperature range of 2K to 10 K. Furthermore, the field dependent specific heat was measured in the same temperature range. The magnetic field was varied from 0T to 7T. Careful subtraction of background was performed by carrying out addenda measurement before starting the experiments.

3. **Results and Discussions**

Figure 1 displays the behavior of experimentally measured magnetic susceptibility as a function of temperature. It can be clearly seen from the plot, that the susceptibility curve shows a rounded maximum at $T_{max}$= 3.7K which is quite close to the earlier reported value [28]. When the temperature is increased beyond 3.7K, the susceptibility gradually decreases. This is indicative of antiferromagnetic linear chain behavior [6]. Previously reported results have satisfactorily established evidence of an isotropic Heisenberg interaction in $Cu(NH_3)_4SO_4 \cdot H_2O$ between the neighboring spins along the chains of $Cu^{++}$ ions [31, 32]. Bonner and Fisher employed the exact diagonalization technique to calculate the magnetic susceptibility for antiferromagnetic spin chain where they varied the number of spins up to 11. However, due the finite size effects, their results cannot satisfactorily explain the experimental results. Subsequently, by following the analytical approach of Bethe ansatz [12], it was possible to solve the one dimensional Heisenberg spin ½ model exactly [33, 34]. Thus, the magnetic susceptibility for uniform spin chain model has been calculated much more accurately down to very low temperature. The numerical data has been efficiently applied to analyze the thermodynamic behavior of certain spin chain materials [26, 27]. The present experimental system is a physical example of uniform isotropic chain model where the neighboring spins of the constituent $Cu^{++}$ ions being arranged in a periodic fashion along one particular direction, interact microscopically. Therefore, the exact susceptibility result derived by Bethe ansatz technique for infinite spin ½ chain has been compared to the experimental data for $Cu(NH_3)_4SO_4 \cdot H_2O$ in the temperature range 1.9K to 30 K. The best match was found for exchange coupling constant J=6K and Landé g factor g=2.056. The theoretical curve (solid red line) appears to be consistent with the experimental data (open circles).

Experimental isothermal magnetization data are taken at low temperature regime such that the antiferromagnetic correlations survive persistently. The magnetization curves taken at 1.9K, 2.5K and 3.4K (which are above the antiferromagnetic ordering temperature 0.37K [32]) are plotted in figure 2 with magnetic field along the horizontal axis. The magnetic field is varied from 0 to 14T whereas at 14T, it can be seen that the magnetization curve at lowest temperature (1.9K) almost reaches its saturation value. ALPS (Algorithms and Libraries for Physics Simulations) provide simulation codes for various strongly correlated quantum mechanical models [35]. Based on the "stochastic series expansion in the directed loop representation" method [36], QMC technique is employed (using code from ALPS) to simulate isothermal magnetization in the same magnetic field range for N=100 spin ½ sites. This analysis was performed for all the three isotherms. These numerically simulated curves are plotted in the same graph with the experimental ones. It is quite evident from the graph that the experimental curves are in well agreement with the corresponding simulated ones (assuming J=6.8K). A small mismatch can be observed between the simulation and experiment both at low and high magnetic field values. This discrepancy can happen due to the fact that the spin chain compound may contain paramagnetic spins due the boundary effect of the chains and other magnetic impurities which also contribute to the magnetization curve, whereas the simulation represents an exact solution for a spin ½ chain with 100 sites. Subsequently, all the experimental magnetization isotherms (from 1.9K to 10K) are used to generate a 3D plot with magnetization, magnetic field and temperature along the three axes. The plot is shown in figure 3. This 3D plot explicitly depicts the variation of magnetization with field and temperature.

Experimentally measured specific heat of $Cu(NH_3)_4SO_4 \cdot H_2O$ single crystals taken from 2K to 10K in zero magnetic field is shown in figure 4 (open circles). The most prominent feature in the data that could be observed is the appearance of a broad maximum at $T_{max}$=3K which matches well with previous results [28, 29]. Subsequently, upon enhancement of the temperature, the specific heat decreases slowly. However, when the temperature was increased further, an upturn in the specific heat curve could be observed which is solely due to the lattice contribution as will be clear from the ensuing analysis. The temperature dependence of specific heat can be represented by the following relation,

$$C(T) = C_m(T) + \beta T^3 \qquad (2)$$

Here $C_m$ is the magnetic specific heat and the lattice contribution is determined by the coefficient $\beta$. We have implemented the Bethe ansatz formulation for specific heat for the case of isotropic Heisenberg chain model to analyze the magnetic part of our experimental specific heat data [33]. To demonstrate the variation of the magnetic part of the measured specific heat with temperature, we have extracted the magnetic specific heat by subtracting the lattice part and plotted in figure 4 (open squares). Subsequently, the exact numerical data calculated by Bethe ansatz technique was compared with the experimental magnetic specific heat in the same plot in figure 4. The best match was obtained for J=6K and $\beta$=0.00218 J Mol$^{-1}$ K$^{-4}$. The obtained value of J is consistent with the previous analysis on magnetic susceptibility data and the estimated value of $\beta$ is close to its reported value [28]. The experimental specific heat data, which comprise contribution from both the magnetic and lattice parts, is compared with the theoretical curve representing the total (magnetic and lattice) contribution which was estimated by using $J=6K$ and $\beta=0.00218$ J Mol$^{-1}$ K$^{-4}$ in equation (2) (shown by the solid blue line). As explained by the well known Debye model, the specific heat of the lattice originates mainly due to the optical phonons at low temperature and varies with temperature as $C_{Debye} = \beta T^3$ which could be derived from the following equation [37],

$$C_{Debye} = 9NK_B \left(\frac{T}{\theta_D}\right)^{T/\theta_D} \int_0^{T/\theta_D} \frac{x^4 e^x}{(e^x-1)^2} dx \qquad (3)$$

Here $\theta_D$ is the Debye temperature. In the low temperature regime, $\beta$ has a simple mathematical relationship with the Debye temperature $\theta_D$.

$$\beta = 12\pi^4 NK_B / 5\theta_D^3 \qquad (4)$$

The estimation of Debye temperature was performed by substituting the value of $\beta$ (0.00218) in the above equation. We got $\theta_D = 96.27K$ for the present compound. The most notable aspect of the specific heat curve is the observed broad peak around 3K which arises due to the intrinsic contribution from the many-level energyspectrum of the Heisenberg spin chain [37]. At very low temperature, the thermal energy is not sufficient to excite the system to excited states, yielding a very low value of specific heat. However, with increase in temperature, the probability of occupation of higher energy states increases and the specific heat starts rising. Subsequently, the rate of absorption of thermal energy reaches its maximum value which is responsible for the broad maximum in the specific heat curve around 3K. Upon further increasing temperature, the specific heat drops down to lower value as the energy levels become populated equally and no differential change in the internal energy occurs.

Figure 5 displays experimental specific heat, C(T), data from 2 to 10K in fields up to 7T. Upon increasing the field, the broad maximum at 3K is suppressed towards lower temperature and has completely disappeared above 5T. This behavior is quite similar to the experimental results obtained in the case of other uniform spin chain materials [26, 27]. Bonner and Fisher numerically estimated specific heat for isotropic Heisenberg case where the number of the spins was varied from 2 to 11 [38]. Subsequently, A Klümper investigated the thermodynamics of infinite spin ½ Heisenberg chain and calculated specific heat as a function of temperature at different applied magnetic fields [39]. They observed that with increasing field, the maxima in the $C_p$ vs. T curve shifts to a lower temperature regime accompanied by a reduction in height which supports our experimental data for the system under investigation. In order to interpret the field dependent specific heat data, we have employed the numerical $C_p$ vs. T datasets (at different applied fields) derived by A Klümper using the Bethe ansatz [39]. The experimental data and the exact solutions (the theoretically generated curves have been scaled by assuming J=6K) for 1T, 3T and 5T are plotted in the same graph. One can conclude that the corresponding experimental and theoretical curves are quite consistent with each other (figure 6). Experimentally measured specific heat vs. temperature datasets at constant fields were used to create the 3D plot shown in figure 7.

Next, the fundamental thermodynamic quantities, namely, internal energy and entropy are quantified for $Cu(NH_3)_4SO_4 \cdot H_2O$ from the experimental specific heat data. In general, the internal energy at some particular temperature T is related to the specific heat by the following equation

$$U(T) = U_{2K} + \int_{2K}^{T} C(T') dT' \tag{5}$$

with $U_{2K}$ being the internal energy at 2K. Numerical integration was carried out on the specific heat data in the temperature range of 2K to 10K and the above mentioned integral equation was used to quantify experimental internal energy for the present compound. Similar analysis was performed for each field dependent specific heat datasets. In order to determine $U_{2K}$, the theoretical treatment of A Klümper is followed where the field evolution of internal energy was investigated for Heisenberg isotropic chain by means of application of Bethe ansatz. We have used J=6K obtained from earlier analyses. Therefore, the theoretical values of internal energies $U_{2K}$ at the particular fields corresponding to the experimental Cp vs. T datasets are substituted in the integral equation (5) to evaluate temperature dependent internal energy datasets [$U(T)$ vs. T] at different fixed fields. Both the theoretical and the experimental energies are scaled in units of Kelvin. These extracted $U(T)$ vs. T datasets are plotted in figure 8. A surface plot has been created using the $U(T)$ vs. T datasets for different fields (figure 9).

Fundamental thermodynamic relations imply that the entropy increment of a system could be simply calculated from specific heat using the relation $\Delta S(T) = S_{2K} + \int_{2K}^{T} [C(T')/T']dT'$. Here $S_{2K}$ is the entropy at 2K. Hence, we substituted the experimental specific heat data in the above equation and integrated numerically to estimate the entropy increment. The above treatment was performed for all C(T) vs. T datasets obtained at fixed fields and these results are shown in figure 10. The integration constant (entropy for S=1/2 Heisenberg spin chain at T=2K) has been determined theoretically [39, 40] and incorporated in the integration. The figure shows that the zero field magnetic entropy saturates at the value of 0.688 which is quite close to the theoretically predicted value of ln(2)=0.693. This observation also supports the fact that the phonon contribution has been efficiently subtracted from the experimental specific heat data. We then used the entropy vs. temperature datasets at different fields to generate a surface plot shown in figure 11. The plot explicitly depicts the behavior of entropy with change in temperature and field. The entropy increases with increase in temperature for all field values as is expected, since antiferromagneic correlations or ordering is destroyed with increase in temperature. At high temperature (~10K), where the antiferromagnetic correlations are minimum, the entropy decreases upon increasing the field owing to some field dependent alignment of the paramagnetic spins.

## 4. Conclusion

The present work exemplifies an investigation of thermodynamics of a Heisenberg spin ½ chain material where various information about the system has been captured. In summary, we have performed a detailed study of thermal and magnetic properties of $Cu(NH_3)_4SO_4·H_2O$ which can be best described by an ideal spin ½ chain with isotropic Heisenberg interaction. The experimental data have been compared with the exact solutions calculated using the Bethe ansatz technique. Temperature dependent susceptibility and specific heat data are in perfect agreement with the calculation using an exchange coupling constant J=6K. This consistency indicates that the material displays spin chain behavior in the temperature range down to 2K. Field dependent magnetization curves were generated numerically using the QMC technique at the Heisenberg extreme. Numerically simulated curves are in agreement with the experimental ones. Temperature dependent specific heat data at various applied fields were also compatible with the exact results for infinite Heisenberg spin chain. Furthermore, numerical integration was performed on the experimental specific heat data to obtain the variation of internal energy and entropy with temperature at various fields.

These spin chain systems can have fruitful applications from the perspective of quantum communications. Bose has described that an entangled spin chain can be used as an appropriate channel for transmitting a quantum state over a short distance [41]. It has been suggested that the above scheme can be efficiently implemented for one-dimensional Heisenberg chain compounds with nearest neighboring interaction where quantum states can be transferred with an improved fidelity compared to classical one [41]. Successful implementation of the above protocol can play a significant role in designing a feasible quantum computer. Moreover, the possibility of having entangled spins in solid state crystals has an advantage over the optical systems as the crystals can be efficiently integrated with existing Si-based technology or other quantum devices [42].

**Acknowledgments**

   The authors would like to thank the Ministry of Human Resource and Development, Government of India and Göttingen Kolkata Open Shell Systems (G-KOSS) for funding. We sincerely acknowledge A. Honecker for useful discussions. The authors are grateful to S. K. Dhar for allowing us to carry out some magnetic measurements in his lab in TIFR, Mumbai. We also acknowledge K. R. Kumar and A. Maurya for their assistance with the magnetic measurements.

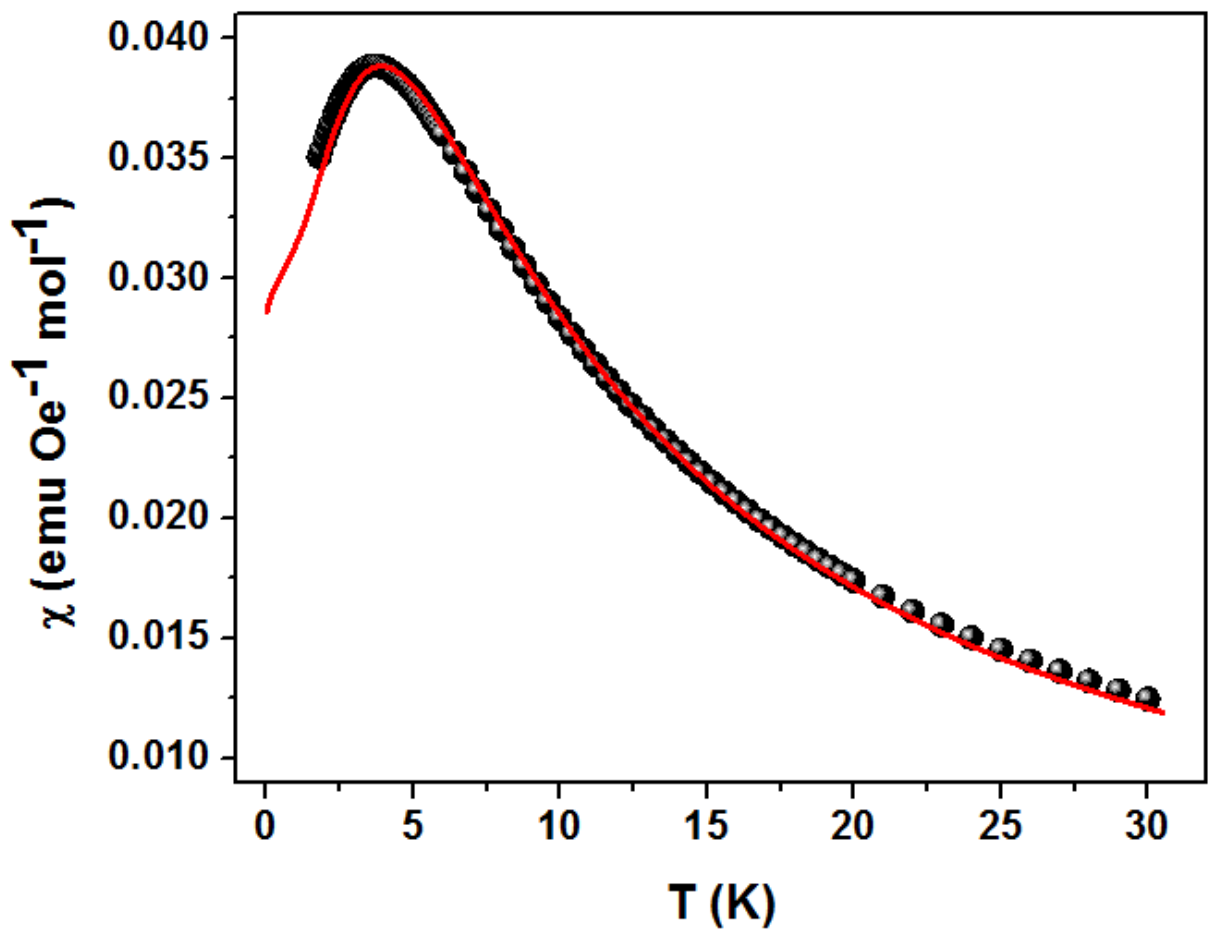

**figure** 1. Magnetic susceptibility versus temperature for $Cu(NH_3)_4SO_4 \cdot H_2O$ (circles represent the experimental data while the exact solution of the S=1/2 Heisenberg model using J =6 and g=2.056 is shown by the solid red curve).

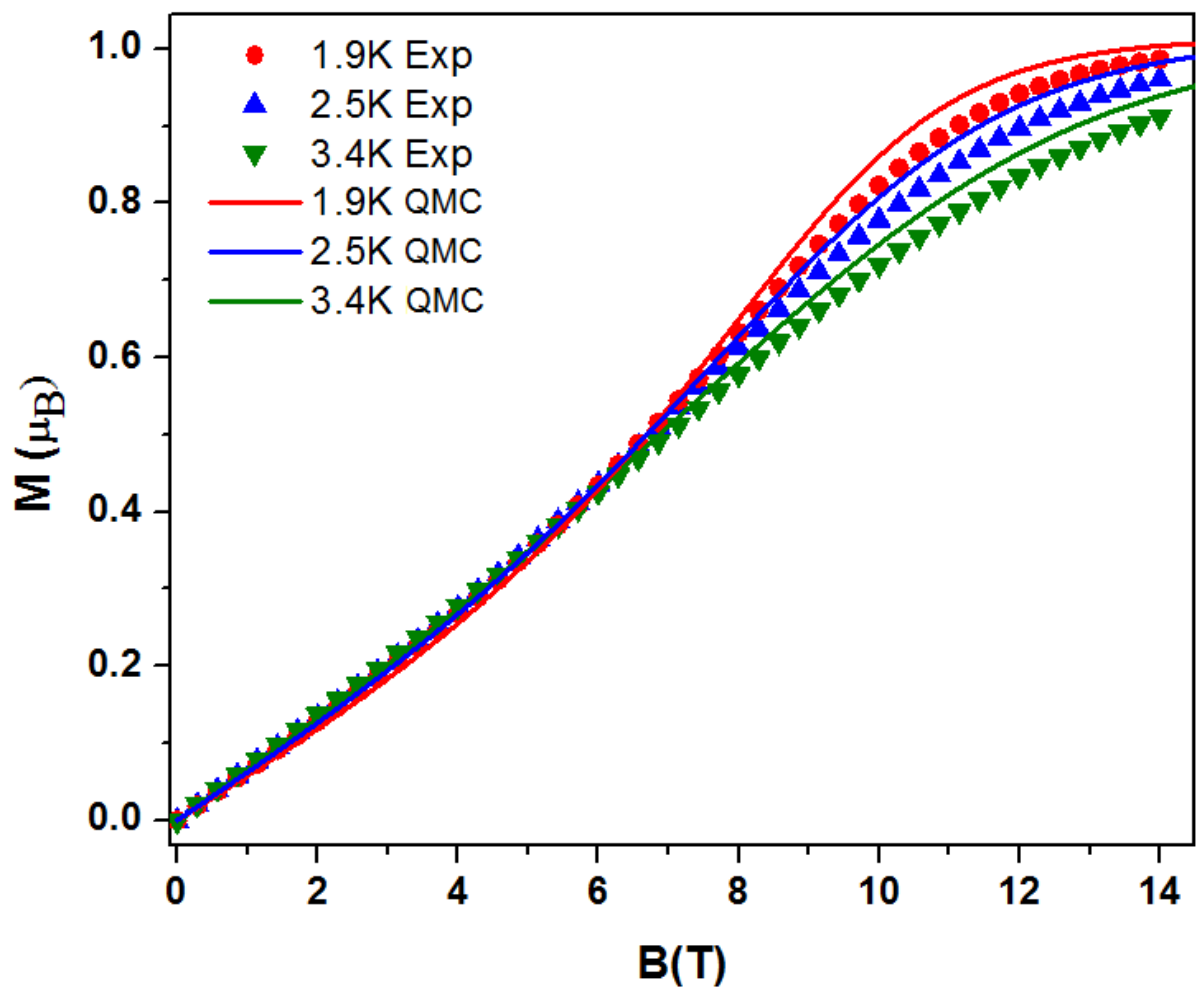

**figure** 2. Experimentally determined magnetization vs. magnetic field at different temperatures (symbols represent data taken at temperatures shown in the legend) along with the corresponding QMC (Quantum Monte Carlo) results derived using the Hamiltonian for a chain consisting of 100 spins.

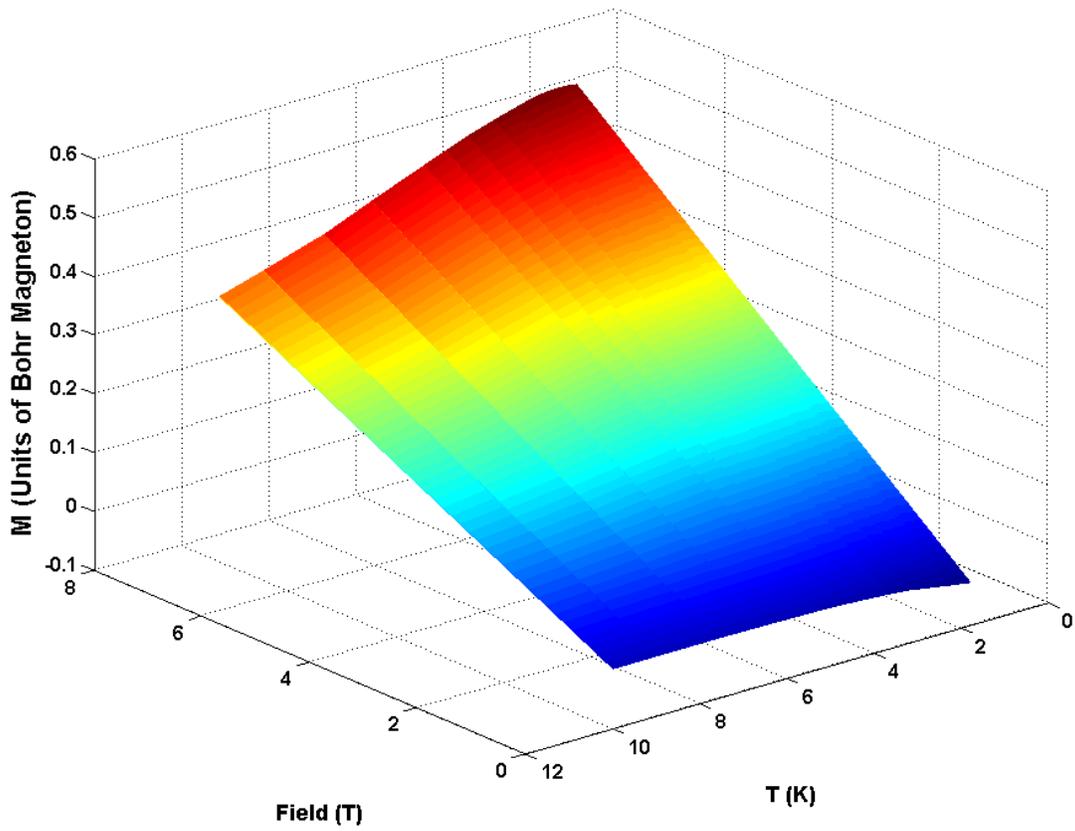

figure 3. Surface plot with magnetization, magnetic field and temperature along the three axes.

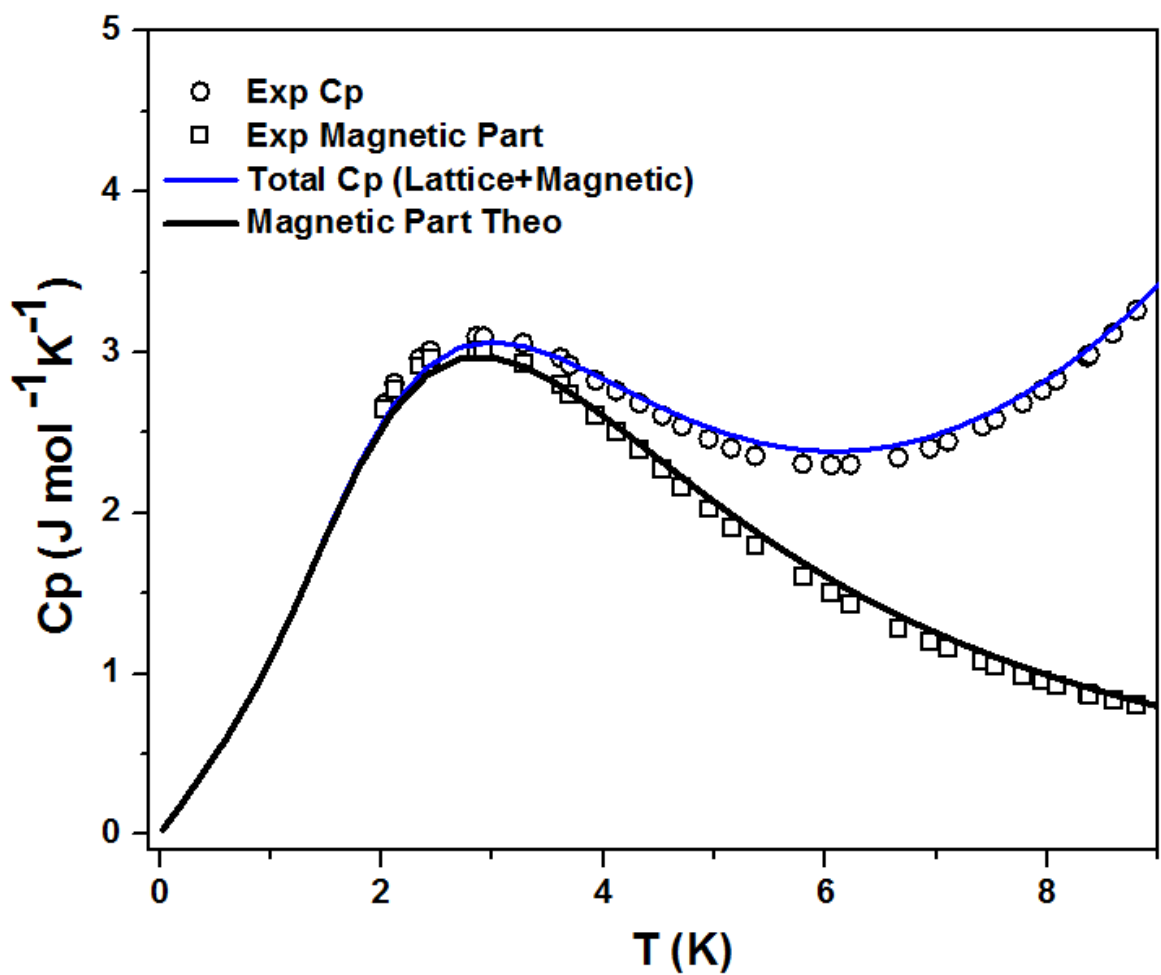

**figure** 4. Total (magnetic and lattice component) experimental specific heat of $Cu(NH_3)_4SO_4 \cdot H_2O$ as indicated by open circles . Solid curves are respective calculated curves as explained in the main text.

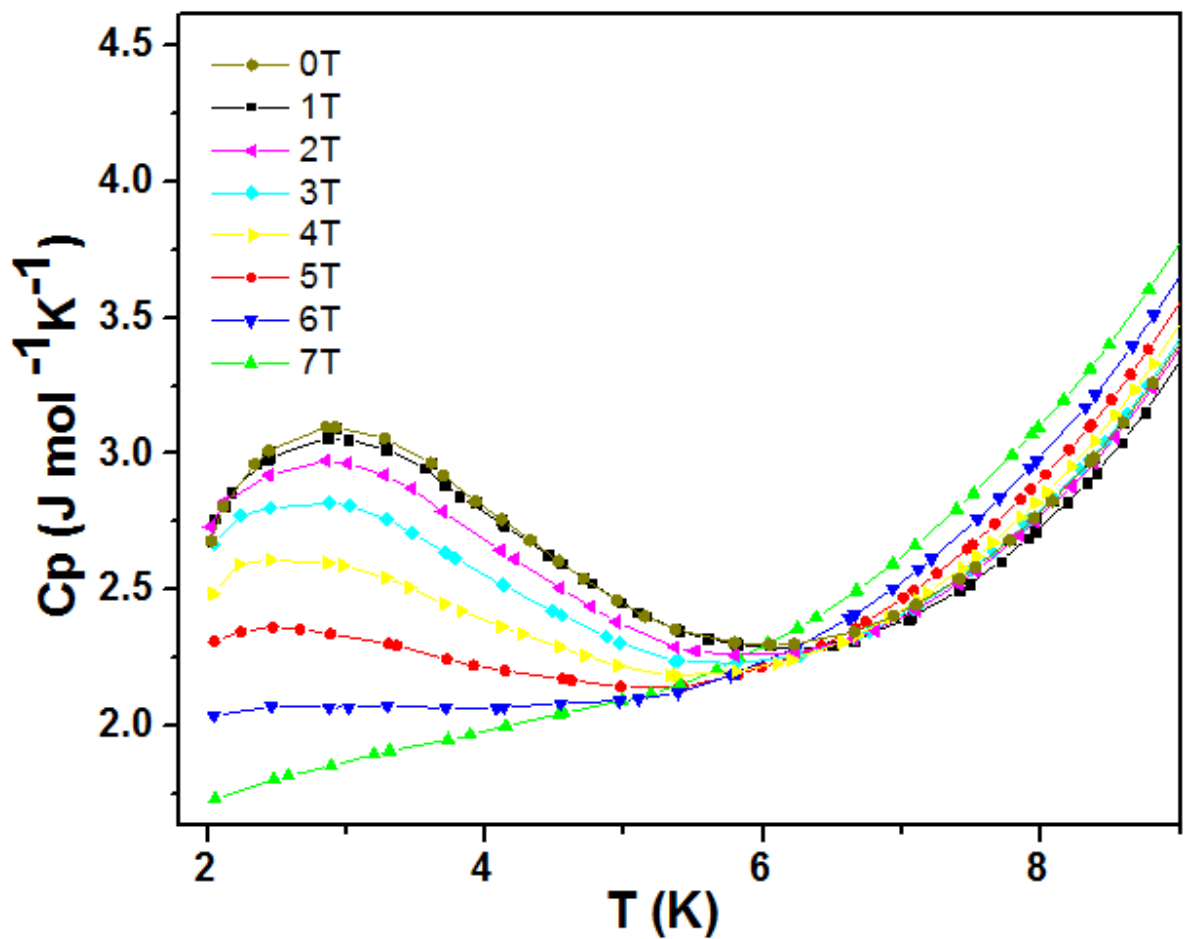

**figure** 5. Experimental specific heat data for Cu(NH$_3$)$_4$SO$_4$. H$_2$O as a function of temperature at different fields.

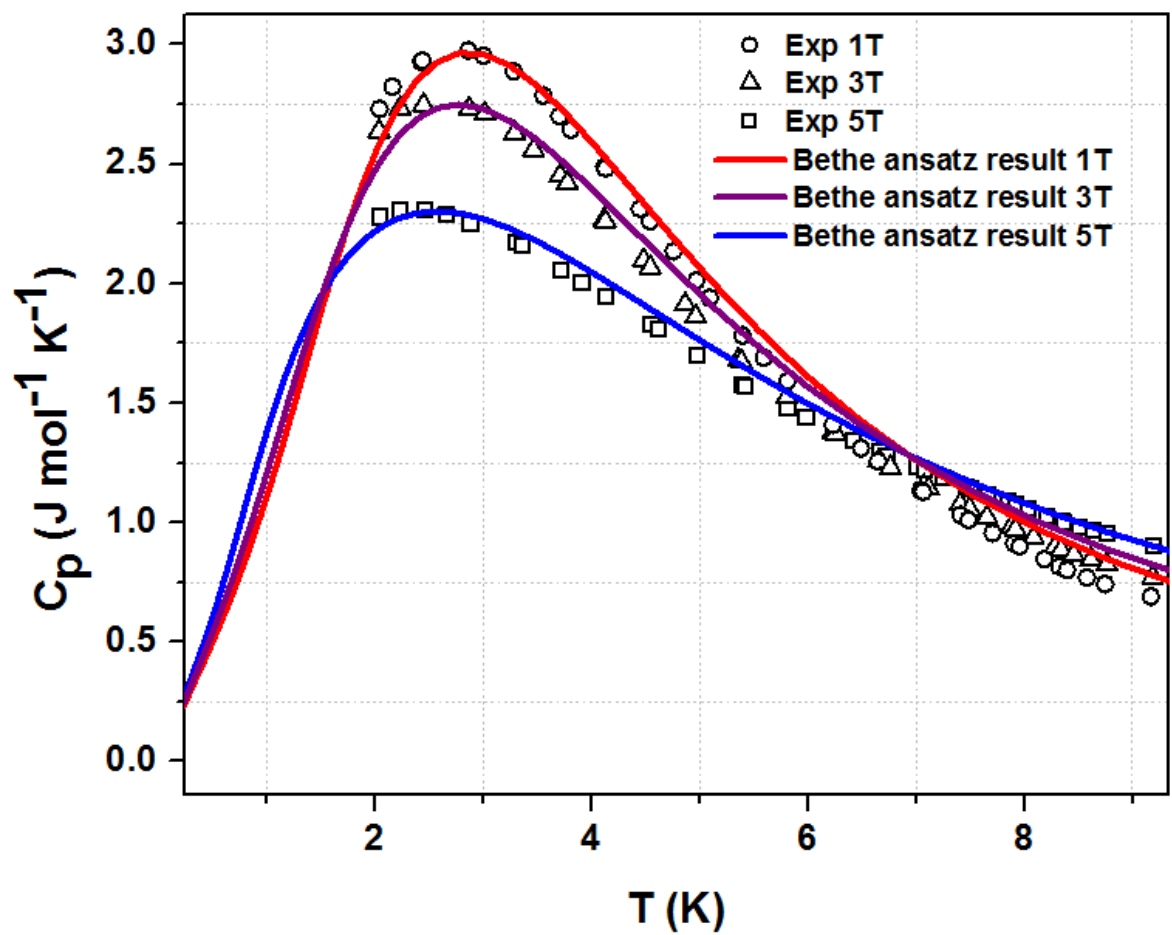

**figure** 6. Comparison of experimental specific heat data with corresponding Bethe ansatz results for magnetic fields of 1T, 3T and 5T.

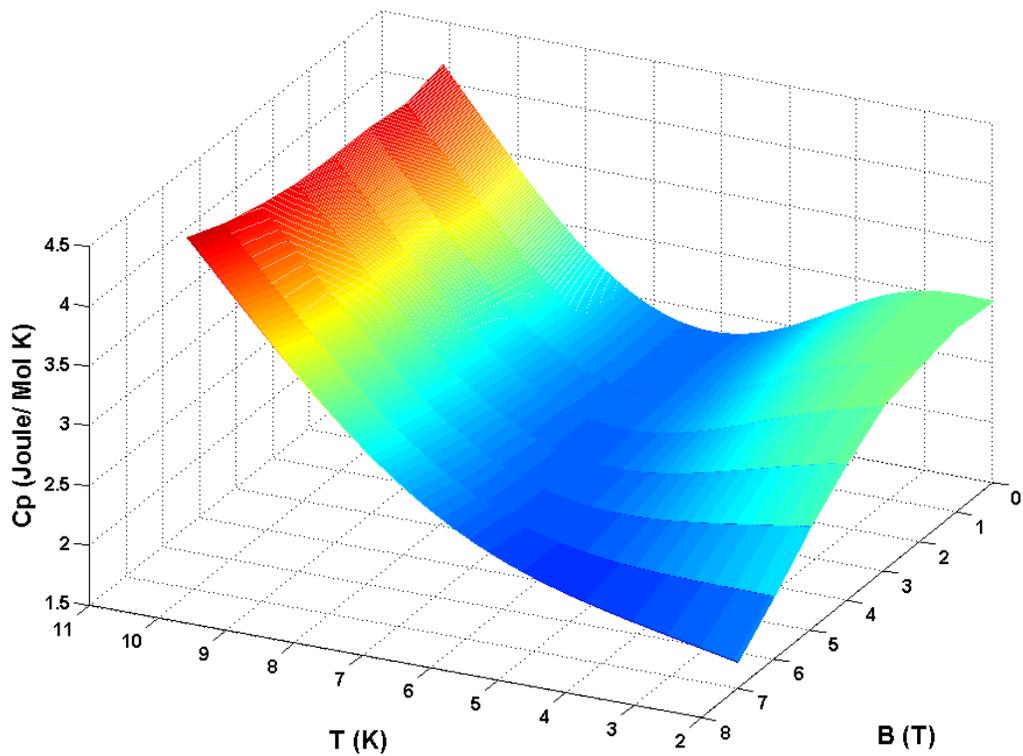

**figure** 7. Three dimensional plot depicting the variation of experimental specific heat with magnetic field and temperature.

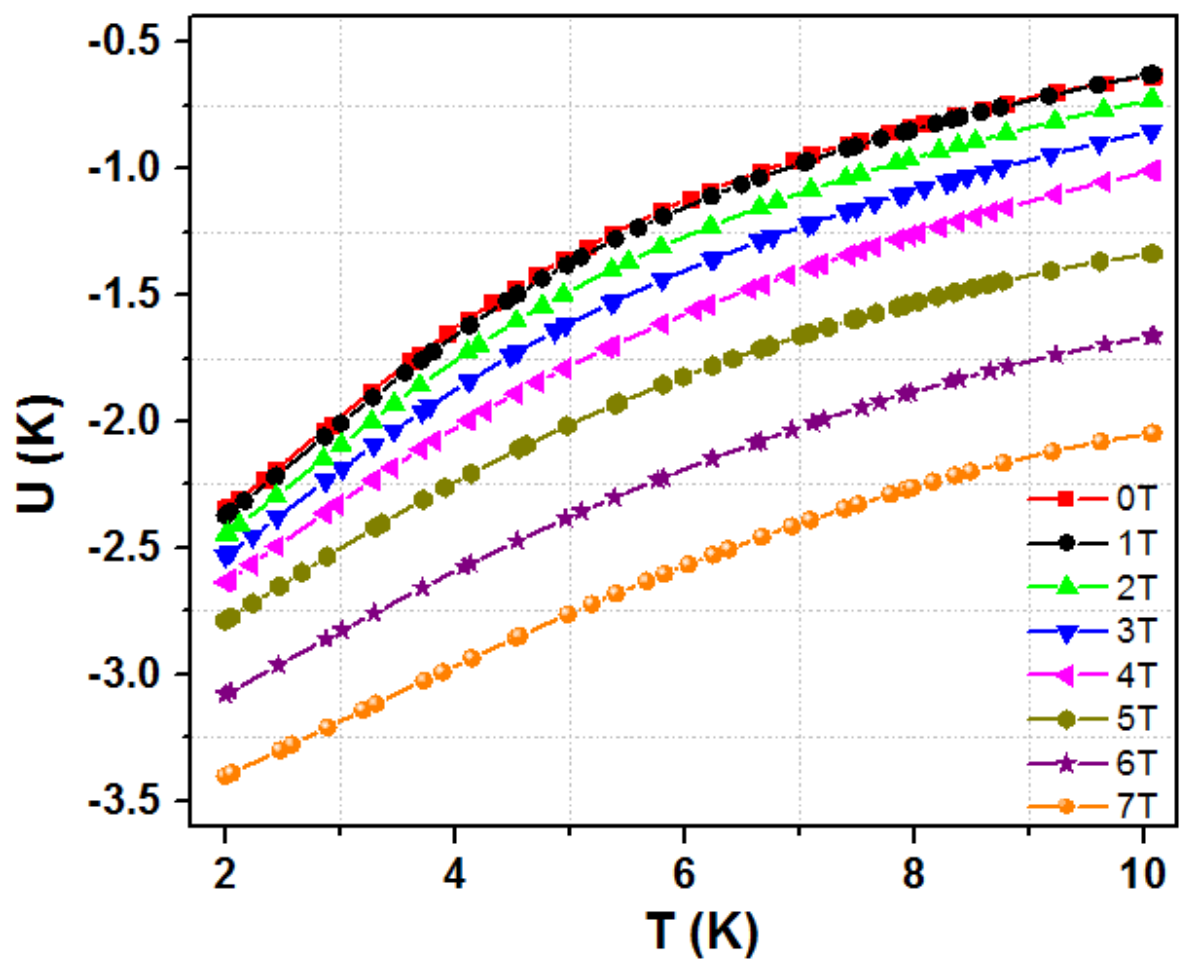

**figure** 8. Variation of experimental internal energy with temperature for different applied magnetic fields (as shown in the legend).

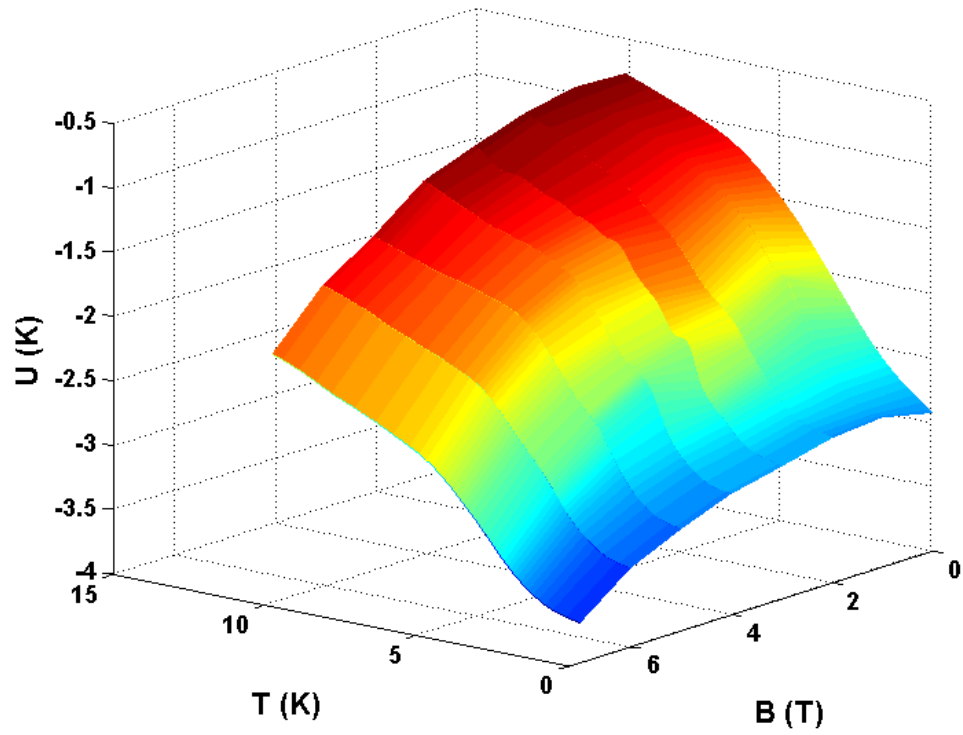

**figure** 9. Three dimensional variation of internal energy with magnetic field and temperature along the other two axes.

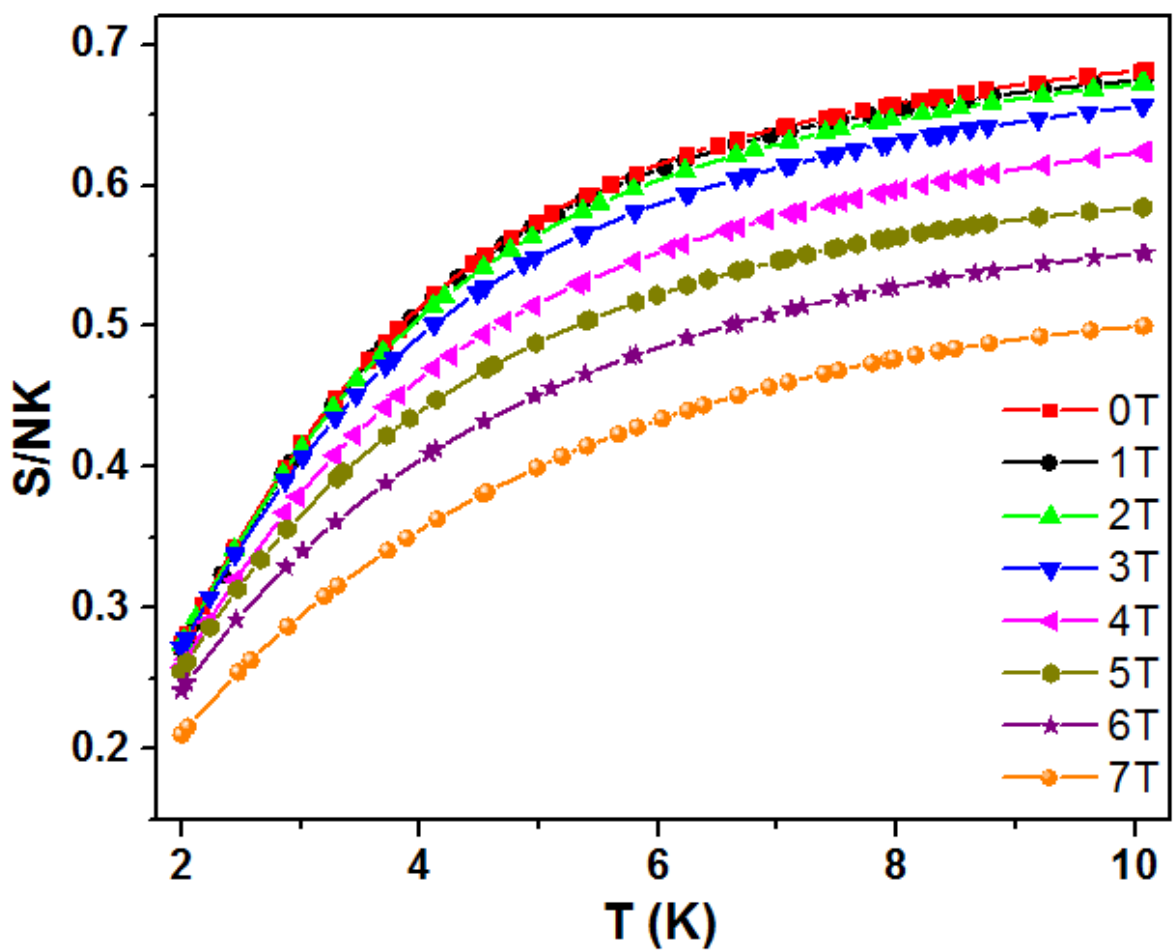

**figure** 10. Temperature dependence of entropy obtained from the experimental specific heat data taken at different fields (as shown in the legend).

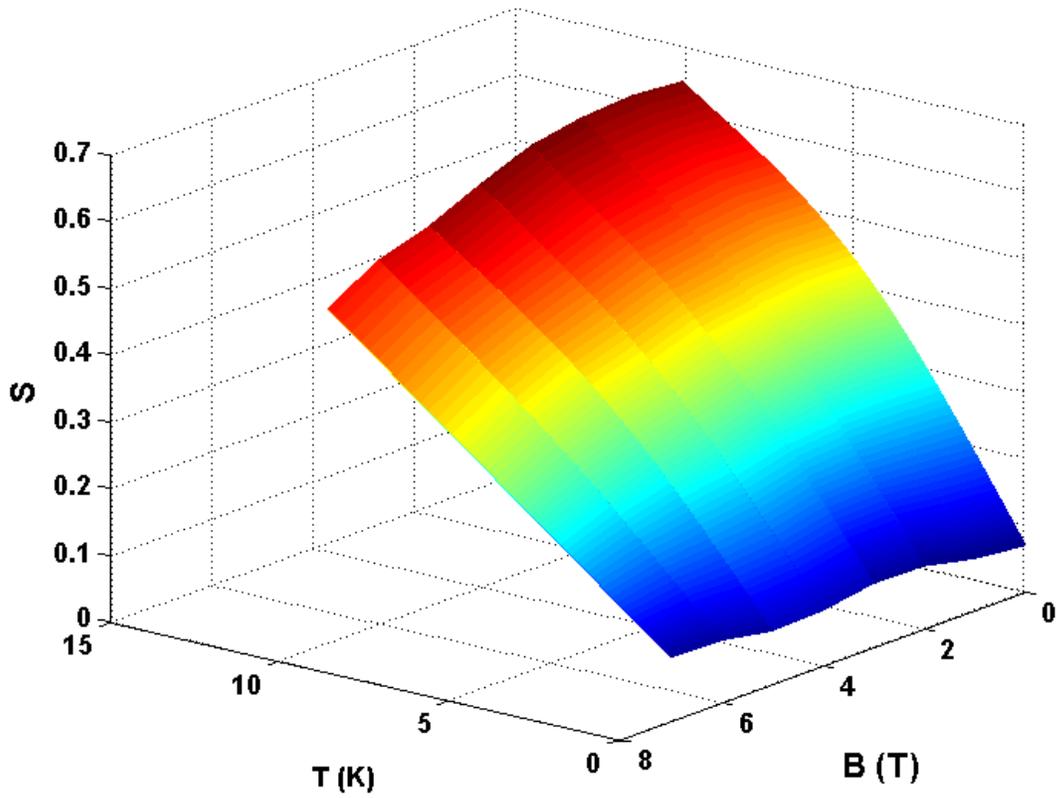

**figure** 11. Three dimensional plot of entropy with magnetic field and temperature.